\documentclass[preprint,showpacs]{revtex4}
\usepackage{graphicx}
\usepackage{amssymb}
\newcommand{\be}{\begin{eqnarray}}
\newcommand{\ee}{\end{eqnarray}}

\begin{document}

\title{Generalized parton distributions for the proton in AdS/QCD}
\author{\bf Dipankar Chakrabarti and Chandan Mondal}
\affiliation{ Department of Physics, 
Indian Institute of Technology Kanpur,
Kanpur 208016, India}
\date{\today}

\begin{abstract}
We present a study of proton generalized parton distributions( GPDs) in both momentum and position spaces using the proton wave function obtained from AdS/ QCD. Here we consider  the soft wall model. The results are compared with  a phenomenological model of proton GPDs.

\end{abstract}
\maketitle


\vskip0.2in
\noindent
{\bf Introduction}

Generalized parton distributions (GPDs) appear in the exclusive processes like deeply virtual Compton scattering (DVCS) or vector meson productions. The GPDs are  
 functions of three variables, namely, longitudinal momentum faction $x$ of the quark or gluon,   square of the total momentum transferred ($t$)  and the skewness $\zeta$, which represents the longitudinal momentum transferred in the process. The GPDs contain much more  informations
about the nucleon structure and spin (see \cite{rev} for example) compared to the ordinary parton distribution functions(pdfs) which are functions of $x$ only.   The GPDs reduce to the ordinary parton distributions in the forward limit.  Their first moments are related to the form factors and 
provide interesting information about the spin and orbital angular momentum of the constituents,  as well as the spatial structure, of the nucleons.  Being off-forward matrix elements, the GPDs have no probabilistic interpretation.
 But for zero skewness, the  Fourier transforms of the GPDs  with respect to the transverse momentum transfer ($\Delta_\perp$) give  the impact parameter dependent GPDs which satisfy the positivity condition and can be interpreted as distribution functions \cite{burk1}. The impact parameter dependent GPDs  provide us with  the information about partonic distributions in the impact parameter or the transverse position space for a given longitudinal momentum ($x$).  The impact parameter $b_\perp$ gives the separation of the struck quark from the center of momentum.  In the $t\to 0$ limit, Ji's sum rule \cite{ji} relates the moment of  the GPDs to the  angular momentum contribution to the nucleon by  the quark or gluon. Again in the impact parameter space, the sum rule has a simple interpretation for a transversely polarized state \cite{burk}. The term containing $E(x,0,0)$ arises due to a transverse deformation of the GPDs in the center of momentum frame, whereas the term containing $H(x,0,0)$ is an overall transverse shift when going from the transversely polarized state in instant form to the front form.
 
 As different experiments have measured or are planned to measure DVCS as well as vector meson production(e.g, HERA H1, COMPASS \cite{H1,H2,COMPASS} and ZEUS  \cite{ZEUS1,ZEUS2} collaborations, HARMES \cite{HERMES}, JLAB \cite{CLAS}, etc.), there are many activities going on to model GPDs for the proton.

In the overlap representation, GPDs can be expressed as off-forward matrix elements of bilocal  light front currents.  Using AdS/QCD, one can extract the light front wave functions(LFWFs) for the hadrons and
   thus provide an interesting way to calculate the GPDs.  Polchinski and Strassler \cite{PS} first used the AdS/CFT duality to address the hard scattering and  deep inelastic scattering. The AdS/QCD for the baryon  has been developed by  several groups 
\cite{BT,ads1,ads2}.
Though it gives only the semiclassical approximation of QCD, so far this method has  been successfully applied to  describe many hadron properties e.g.,   the hadron mass spectrum, parton distribution functions, meson and nucleon form factors,  structure functions etc\cite{BT1,AC,BT2}.   To describe  the Pauli form factor in AdS/QCD, one needs to have nonminimal coupling as proposed by Abidin and Carlson \cite{AC}. 
Recently it has been shown that AdS/QCD wave functions remarkably agree with experimental data for $\rho$ meson electroproduction \cite{forshaw}. 
AdS/QCD has also been successfully applied in the meson sector to predict the branching ratio for decays of $\bar{B^0}$ and $\bar{B_s^0}$ into $\rho$ mesons \cite{ahmady1}, isospin asymmetry and branching ratio for the $B\to K^*\gamma$ decays \cite{ahmady2}, etc.  An interesting study of Ehrenfest correspondence principle and the AdS/QFT duality has been done in \cite{glazek}.
Studies of the nucleon form factors with higher Fock  sectors have been done in \cite{hf1}.  Vega {\it et al.} \cite{vega} proposed a beautiful prescription to extract GPDs from the form factors in AdS/QCD and they have done the  GPD calculations using both the hard and soft wall models in AdS/QCD.  It has been shown that the form of the confining potential  in the soft wall model  is unique for both the meson and baryon sectors \cite{BTD}.
In this work, we will provide the results for GPDs using the LFWFs obtained from the AdS/ QCD. We will use the formula for the nucleon form factors in the light front quark model with SU(6) spin flavor symmetry and  compare the GPDs in  the impact parameter space with a phenomenological model of  the GPDs for the proton.

\vskip0.2in
\noindent
{\bf  Baryon wave functions from AdS/QCD}

In this section we briefly review the derivation of the baryon wave functions in AdS/QCD following Brodsky and T\'eramond \cite{BT,BT2}.
We know that the AdS/CFT correspondence relates a gravitationally interacting theory in anti de Sitter space $AdS_{d+1} $ with a conformal gauge  theory in $d$ dimensions residing at the boundary. Since QCD is not a conformal theory,  one needs to break the conformal invariance of the above duality to generate a bound state spectrum and to relate with QCD. There are two models in the literature to do so.
One is the hard wall model in which the conformal symmetry is broken by introducing a boundary at $z_0\sim1/\Lambda_{QCD}$  in the AdS direction where the wave function is made to vanish. In the soft wall model, the conformal invariance is broken by introducing a confining potential in the action of a Dirac field propagating in $AdS_{d+1}$ space. 
 We will consider the soft model in this paper. The relevant action in the soft model  is written as\cite{BT2}
\be
S=\int d^4x dz \sqrt{g}\Big( \frac{i}{2}\bar\Psi e^M_A\Gamma^AD_M\Psi -\frac{i}{2}(D_M\bar{\Psi})e^M_A\Gamma^A\Psi-\mu\bar{\Psi}\Psi-V(z)\bar{\Psi}\Psi\Big),
\ee
where $e^M_A=(z/R)\delta^M_A$ is the inverse  vielbein and $V(z)$ is the confining potential,
 $R$ is the AdS radius. The  corresponding Dirac equation in AdS is given by
\be
i\Big(z \eta^{MN}\Gamma_M\partial_N+\frac{d}{2}\Gamma_z\Big)\Psi -\mu R\Psi-RV(z)\Psi=0.\label{ads_DE}
\ee
With $z$ identified as the  light front transverse impact variable $\zeta$ which gives the separation of the  quark and 
gluonic constituents in the hadron, it is possible to extract  the light front wave functions for the hadron. 
In $d=4$ dimensions, $\Gamma_A=\{\gamma_\mu, -i\gamma_5\}$.
The light front  wave equation for a baryon  in $2\times 2$ spinor representation can be written as
\be
\frac{d}{d\zeta}\psi_+-\frac{\nu+1/2}{\zeta}\psi_++U(\zeta)\psi_+=\cal{M}\psi_-,\label{lf_DE1}\\
-\frac{d}{d\zeta}\psi_--\frac{\nu+1/2}{\zeta}\psi_-+U(\zeta)\psi_-=\cal{M}\psi_+,\label{lf_DE2}
\ee
where $\nu$ is related to the orbital angular momentum by $\nu=L+1$,  $U(\zeta)=(R/\zeta) V(\zeta)$ is the effective confining potential  in the light front Dirac equation, and $\zeta$ is the light front transverse variable  giving the separation of quark and gluonic constituents in the baryon.
With $z\to\zeta$,   substituting $\Psi(x,\zeta)=e^{-iP\cdot x}\zeta^2\psi(\zeta)u(P)$ in Eq.(\ref{ads_DE}),  and identifying $\mid \mu R\mid=\nu+1/2$,  we  arrive at the Eqs(\ref{lf_DE1}), and (\ref{lf_DE2}).
 For linear confining potential  $U(\zeta)=\kappa^2\zeta$,  Eqs. (\ref{lf_DE1}) and (,\ref{lf_DE2}) can be written as
 \be
 \big(-\frac{d^2}{d\zeta^2}-\frac{1-4\nu^2}{4\zeta^2}+\kappa^4\zeta^2+2(\nu+1)\kappa^2\Big)\psi_+(\zeta)&=&{\cal{M}}^2\psi_+(\zeta)\\
 \big(-\frac{d^2}{d\zeta^2}-\frac{1-4(\nu+1)^2}{4\zeta^2}+\kappa^4\zeta^2+2\nu\kappa^2\Big)\psi_-(\zeta)&=&{\cal{M}}^2\psi_-(\zeta).
 \ee
 
  In case of mesons, the similar potential $\kappa^4\zeta^2$ appears in the Klein-Gordon equation  which can be generated by introducing a dilaton background $\phi=e^{\pm\kappa^2 z^2}$ in the AdS space which breaks the conformal invariance. But in the case of baryons, the dilaton can be scaled out by a field redefinition\cite{BT2}. So, the confining potential for baryons cannot be produced by  the dilaton and is put in by hand in the soft wall model.
  The solutions of the above equations are
  \be
  \psi_+(\zeta)&\sim & \zeta^{\nu+1/2} e^{-\kappa^2\zeta^2/2} L_n^\nu(\kappa^2\zeta^2)\\
   \psi_-(\zeta) &\sim & \zeta^{\nu+3/2} e^{-\kappa^2\zeta^2/2} L_n^{\nu+1}(\kappa^2\zeta^2),
   \ee
with eigenvalue ${\cal{M} }^2=4\kappa^2(n+\nu+1)$. So, the linear confining potential generates a mass gap of the order $\kappa$.
\vskip0.2in
\noindent
{\bf   GPDs from Form Factors} 

The Dirac and Pauli form factors for the nucleons are given by\cite{diehl}
\be
F_1^p(t) &=& \int_0^1 dx(\frac{2}{3} H_v^u(x,t)-\frac{1}{3}H_v^d(x,t)), \nonumber\\
F_1^n(t) &=& \int_0^1 dx(\frac{2}{3} H_v^d(x,t)-\frac{1}{3}H_v^u(x,t)), \nonumber\\
F_2^p(t) &=& \int_0^1 dx(\frac{2}{3} E_v^u(x,t)-\frac{1}{3}E_v^d(x,t)), \label{FF}\\
F_2^n(t) &=& \int_0^1 dx(\frac{2}{3} E_v^d(x,t)-\frac{1}{3}E_v^u(x,t)). \nonumber 
\ee
 Here $x$ is the fraction of the light
cone momentum carried by the active quark and the GPDs for valence quark $q$ are  defined as $H_v^q(x,t)=H^q(x,0,t)+H^q(-x,0,t); ~ E_v^q(x,t)=E^q(x,0,t)+E^q(-x,0,t). $ The GPDs at $-x$  for quarks is equal to the GPDs at $x$ for antiquarks with a minus sign. 

From the AdS/QCD action, the spin nonflip form factors can be written as
\be
F_{\pm}(Q^2)=g_{\pm} R^4\int \frac{dz}{z^4} V(Q^2,z)\mid\psi_{\pm}(z)\mid^2,\label{adsF1}
\ee
where, the coefficients $g_\pm$ are determined from the spin-flavor structure of the model.
The SU(6) spin-flavor symmetric quark model is constructed in the AdS/QCD by weighing the different Fock-state components by the charges and spin projections of the partons as dictated by the symmetry.  In the model, the probabilities to find a quark $q$ in proton or neutron with spin up or down are given by\cite{BT2}
$
N_{p\uparrow}^u=\frac{5}{3},~ N_{p\downarrow}^u=\frac{1}{3}, ~N_{p\uparrow}^d=\frac{1}{3}, N_{p\downarrow}^d=\frac{2}{3},~
N_{n\uparrow}^u=\frac{1}{3},~ N_{n\downarrow}^u=\frac{2}{3}, ~N_{n\uparrow}^d=\frac{5}{3}, N_{n\downarrow}^d=\frac{1}{3}.
$
The coefficients $g_\pm$ in Eq.(\ref{adsF1} ) for a proton and neutron  are then 
$
g_p^+=N_{p\uparrow}^u e_u+N_{p\uparrow}^d e_d=1,~~ g_p^-=N_{p\downarrow}^u e_u+N_{p\downarrow}^d e_d=0,~
g_n^+=N_{n\uparrow}^u e_u+N_{n\uparrow}^d e_d=-\frac{1}{3},~~ g_n^-=N_{n\downarrow}^u e_u+N_{n\downarrow}^d e_d=\frac{1}{3}.
$
In terms of the LFWF derived from the AdS/QCD, the Dirac form factors for the nucleons  in this model are given by \cite{BT2}
\be
F_1^p(Q^2)&=&R^4\int \frac{dz}{z^4} V(Q^2,z)\psi^2_+(z),\\
F_1^n(Q^2) &=& -\frac{1}{3}R^4\int \frac{dz}{z^4} V(q^2,z)(\psi^2_+(z)-\psi^2_-(z)),
\ee
with the normalization conditions $F_1^{p/n}(0)=e_{p/n}$,  the electric charge of the nucleon. 
For Pauli form factors, one needs to include a nonminimal electromagnetic interaction term as proposed by Abidin and Carlson\cite{AC}
\be
\int d^4x ~dz ~\sqrt{g}~\bar{\Psi}~ e^A_M e^B_N [\Gamma_A,\Gamma_B]\Psi.
\ee
This additional term produces the Pauli form factors as
\be
F_2^{p/n}(Q^2) \sim \int \frac{dz}{z^3}\psi_+(z) V(Q^2,z)\psi_-(z).
\ee
The bulk-to-boundary propagator for the  soft wall model
is given by
\be
 V(Q^2,z)=\Gamma(1+\frac{Q^2}{4\kappa^2})U\big(\frac{Q^2}{4\kappa^2},0,\kappa^2 z^2),
 \ee
 where $U(a,b,z)$ is the Tricomi confluent hypergeometric function given by
 \be
 \Gamma(a)U(a,b,z)=\int_0^\infty e^{-zx} x^{a-1}(1+x)^{b-a-1}dx.
 \ee
The above propagator can be written in a simple integral form \cite{Rad, BT2}:
\be
 V(Q^2,z)=\kappa^2z^2\int_0^1\frac{dx}{(1-x)^2} x^{Q^2/(4\kappa^2)} e^{-\kappa^2 z^2 x/(1-x)}.
 \ee
 
\begin{figure}[htbp]
\small{(a)}\includegraphics[width=7.5cm,height=7.5cm,clip]{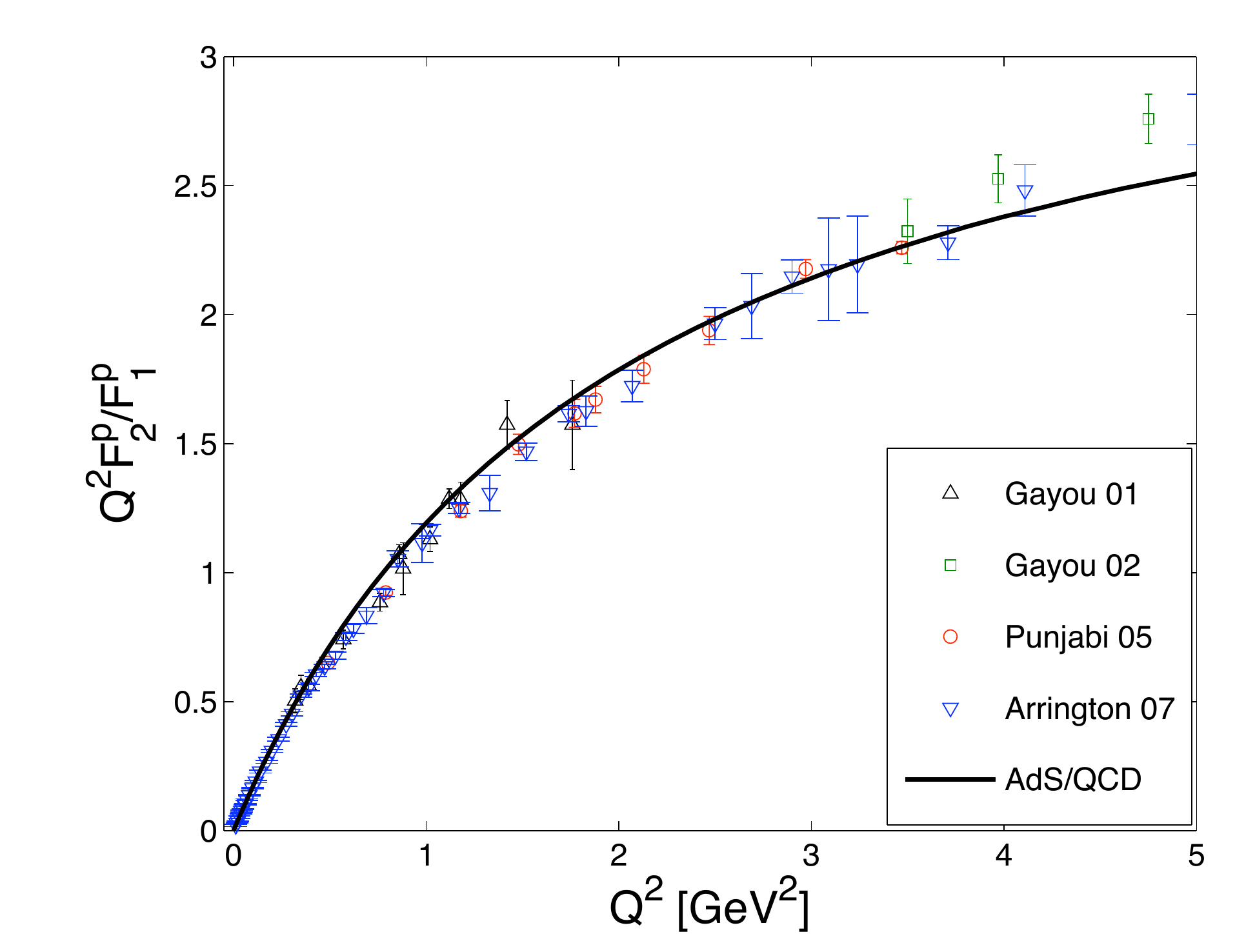}
\small{(b)}\includegraphics[width=7.5cm,height=7.5cm,clip]{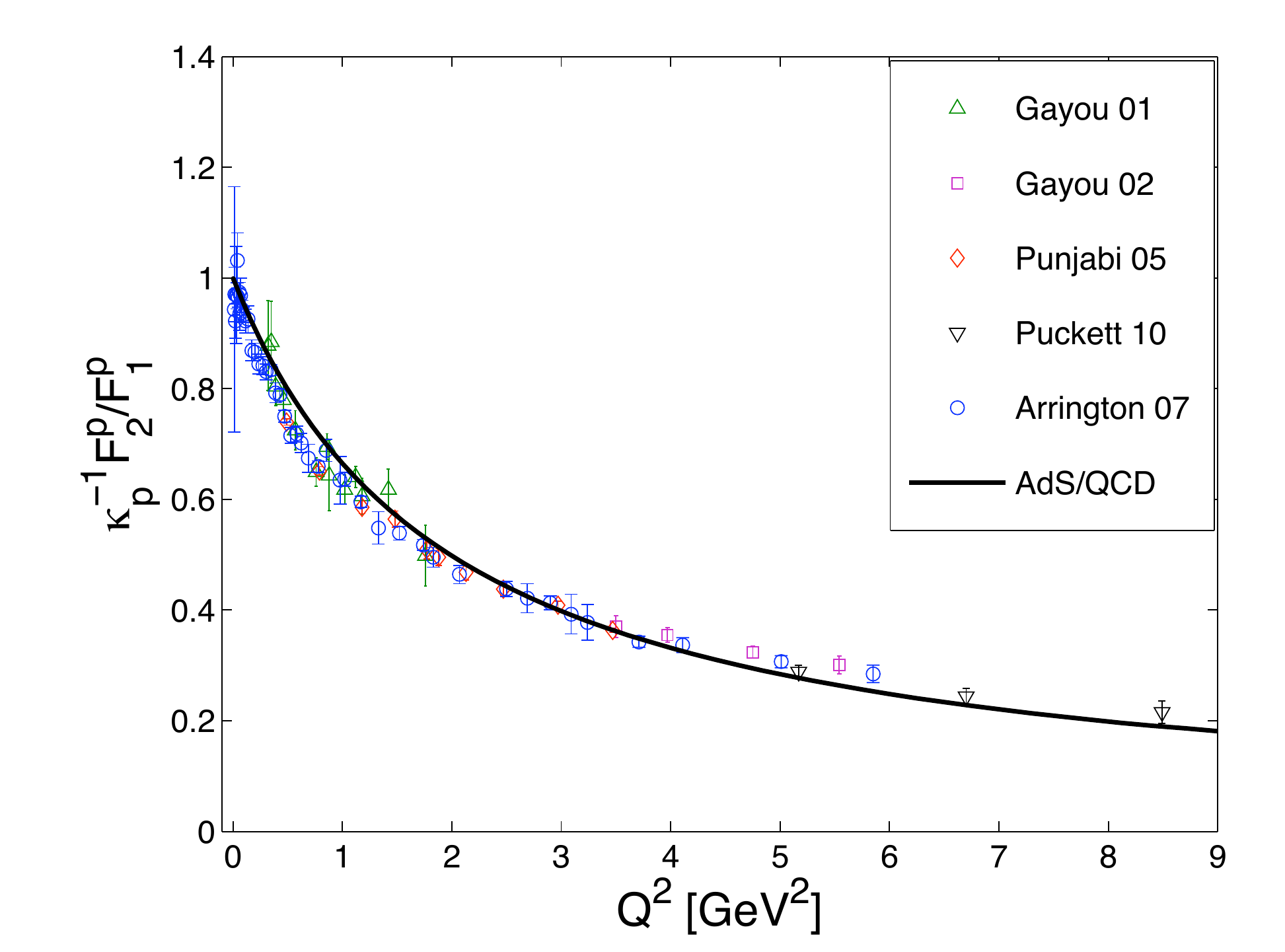}
\caption{\label{fit}(Color online)  AdS/QCD results are fitted with the experimental data.
The plots show the ratio of Pauli and Dirac form factors for the proton, (a) the ratio is multiplied by $Q^2=-q^2=-t$, (b) the ratio is divided by $\kappa_p$. The experimental  data are taken from Refs.  \cite{Gay1,Gay2,Arr,Pun,Puck}.}
\end{figure}

The twist-3 nucleon wave functions in the soft wall model are obtained as  
\be
\psi_+(z) &= &\frac{\sqrt{2}\kappa^2}{R^2}z^{7/2} e^{-\kappa^2 z^2/2},\\
\psi_-(z) &= & \frac{\kappa^3}{R^2}z^{9/2} e^{-\kappa^2 z^2/2}.
\ee
With these wave functions the Pauli form factors fitted to the static values $F_2^{p/n}(0)=\chi_{p/n}$ where $\chi_{p/n}$ is the anomalous magnetic moment of the proton/neutron are written as
\be
 F_2^{p/n}(Q^2) =\kappa_{p/n} R^4 \int \frac{dz}{z^4} V(Q^2,z)\psi_-^2(z).
\ee
We  use the  integral form of the bulk-to-boundary propagator in the formulas for the form factors in AdS space to extract the GPDs using the formulas in Eq. (\ref{FF}).  Here we cannot use the form of form factors in terms of the product of poles on the Regge trajectory as written in \cite{BT2}, but we need to use the explicit    formulas of the form factors as stated above with the wave functions so that we can exploit the integral form of the bulk-to-boundary propagator to extract the GPDs. We find that the best fit to the form factors is obtained for $\kappa=0.4066$ GeV as shown in Fig.\ref{fit}.

\begin{figure}[htbp]
\begin{minipage}[c]{0.98\textwidth}
\small{(a)}
\includegraphics[width=7.5cm,height=7.5cm,clip]{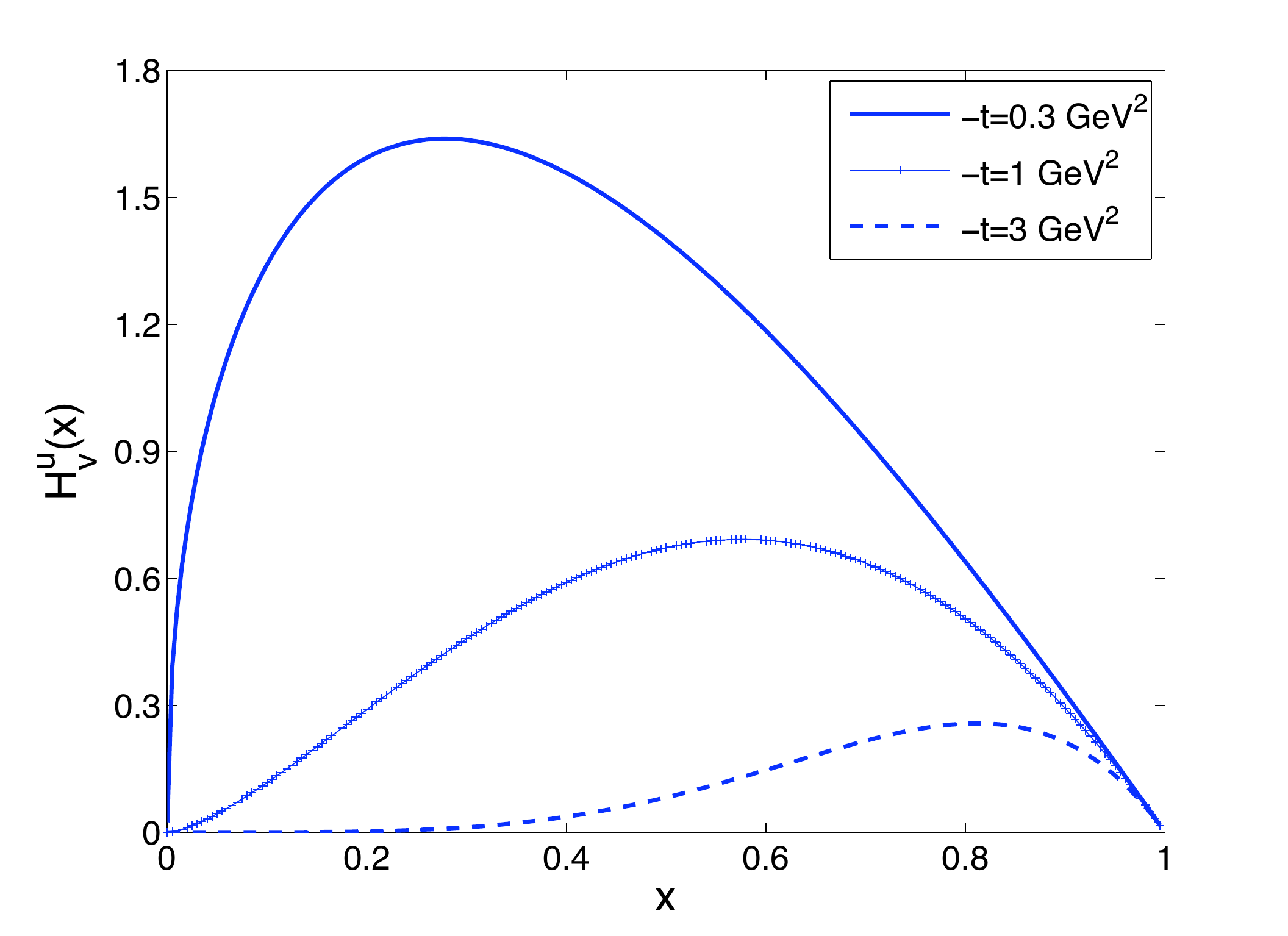}
\hspace{0.1cm}%
\small{(b)}\includegraphics[width=7.5cm,height=7.5cm,clip]{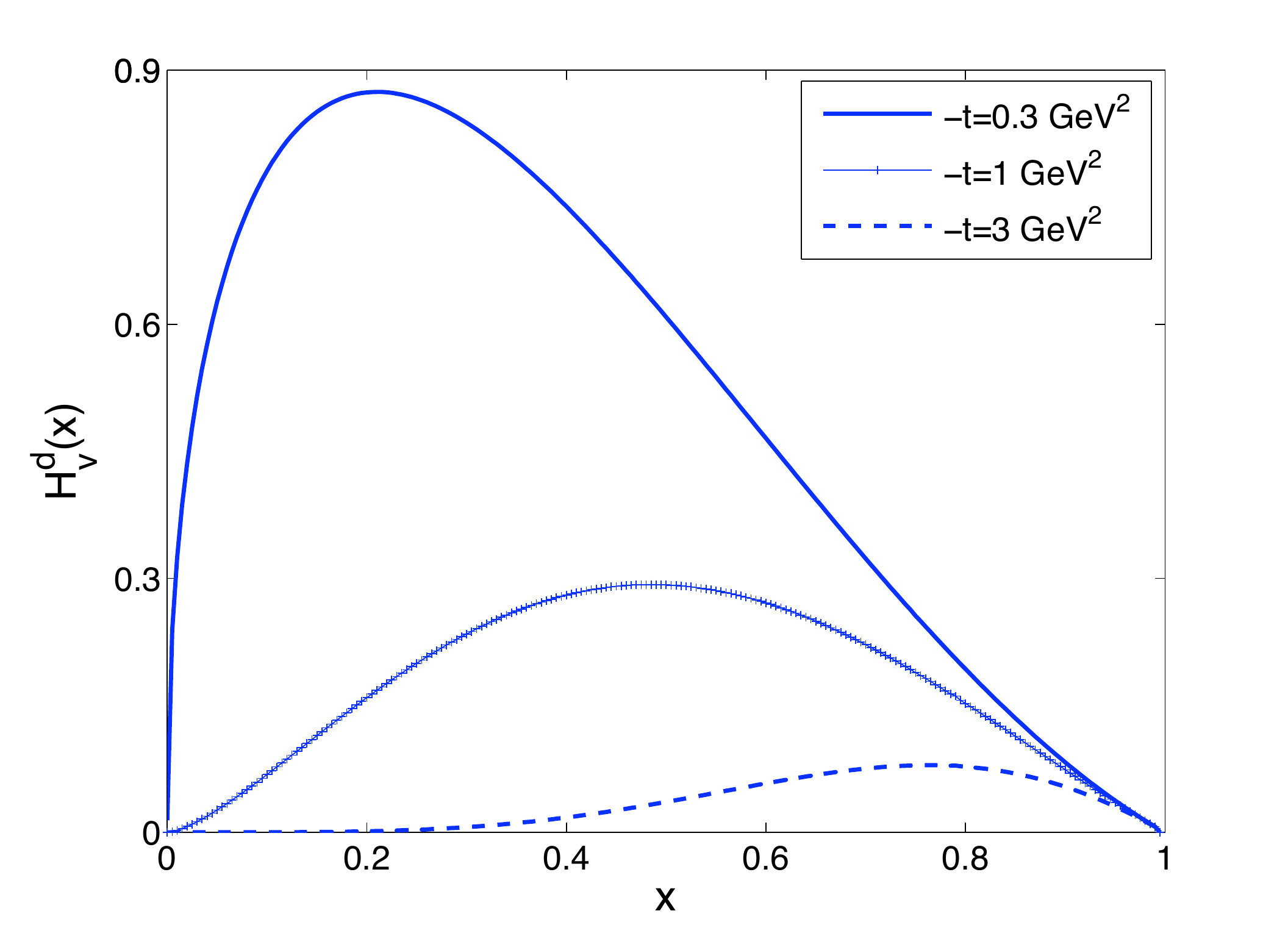}
\end{minipage}
\caption{\label{Hv_x}(Color online) Plots of (a) $H^u(x,t)$ vs $x$ for fixed values of $-t$    (b) the same as in (a) but for $d$ quark.}
\end{figure}

\begin{figure}[htbp]
\begin{minipage}[c]{0.98\textwidth}
\small{(a)}
\includegraphics[width=7.5cm,height=7.5cm,clip]{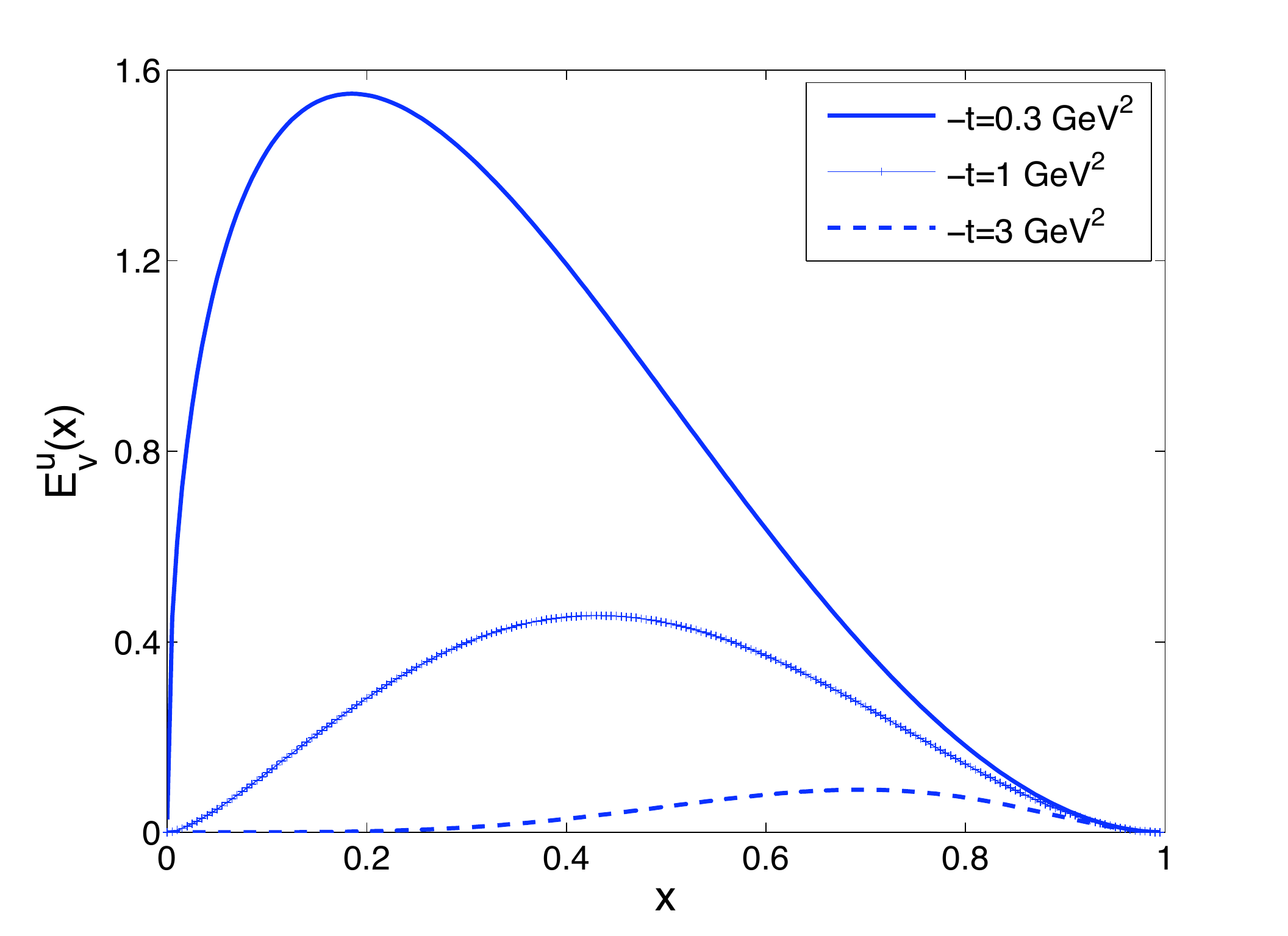}
\hspace{0.1cm}%
\small{(b)}\includegraphics[width=7.5cm,height=7.5cm,clip]{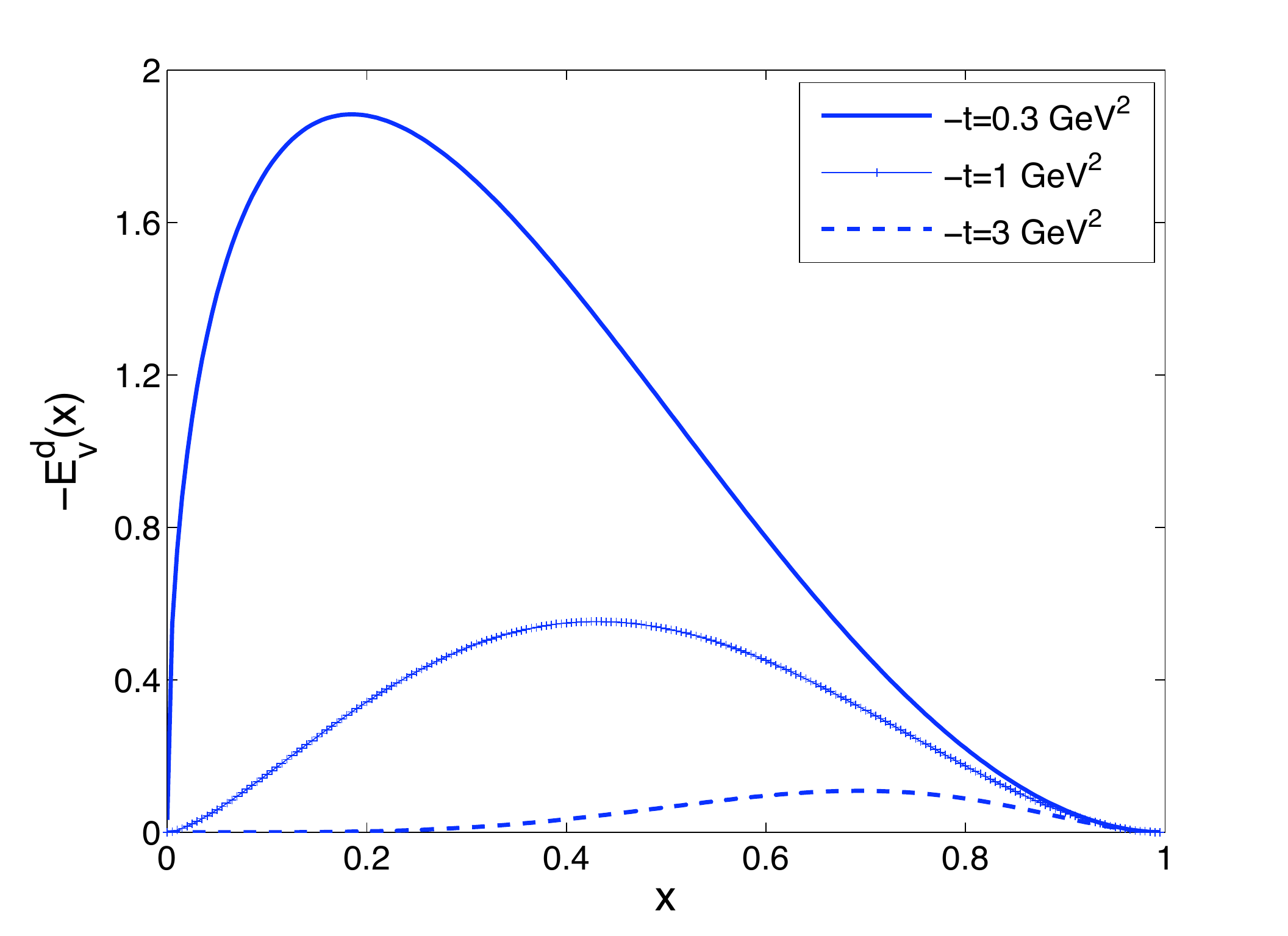}
\end{minipage}
\caption{\label{Ev_x}(Color online) Plots of (a) $E^u(x,t)$ vs $x$ for fixed values of $-t$;    (b) $-E^d(x,t)$ vs $x$ for fixed values of $-t$.}
\end{figure}

In the Figs.\ref{Hv_x} (a) and (b)  we  show the GPD $H(x,t)$ as functions of $x$ for different $-t$ values for u and d quarks.
Except for the fact that it falls off faster as $x$ increases  for d quarks, the overall nature is the same for both u and d quarks.
The apparent similarity in the behaviors for $u$ and $d$ quarks  might be due to the $SU(6)$ symmetry in the AdS/QCD model.
Similarly in Figs.\ref{Ev_x}(a) and (b) we  show the GPD $E(x,t)$ as a function of $x$ for different $-t$ for $u$ and $d$ quarks. Unlike $H(x,t)$, the falloff of the GPD $E(x,t)$ with increasing $x$ is similar for both $u$ and $d$ quarks.

\vskip0.2in
\noindent
{\bf GPDs in impact parameter space}

GPDs  in transverse impact parameter space are defined
as \cite{burkardt}:
\be
H(x,b)={1\over (2 \pi)^2} \int d^2 \Delta e^{-i \Delta^\perp \cdot b^\perp}
H(x,t),\nonumber\\
E(x,b)={1\over (2 \pi)^2} \int d^2 \Delta e^{-i \Delta^\perp \cdot b^\perp}
E(x, t).
\ee

\begin{figure}[htbp]
\begin{minipage}[c]{0.98\textwidth}
\small{(a)}
\includegraphics[width=7cm,height=6cm,clip]{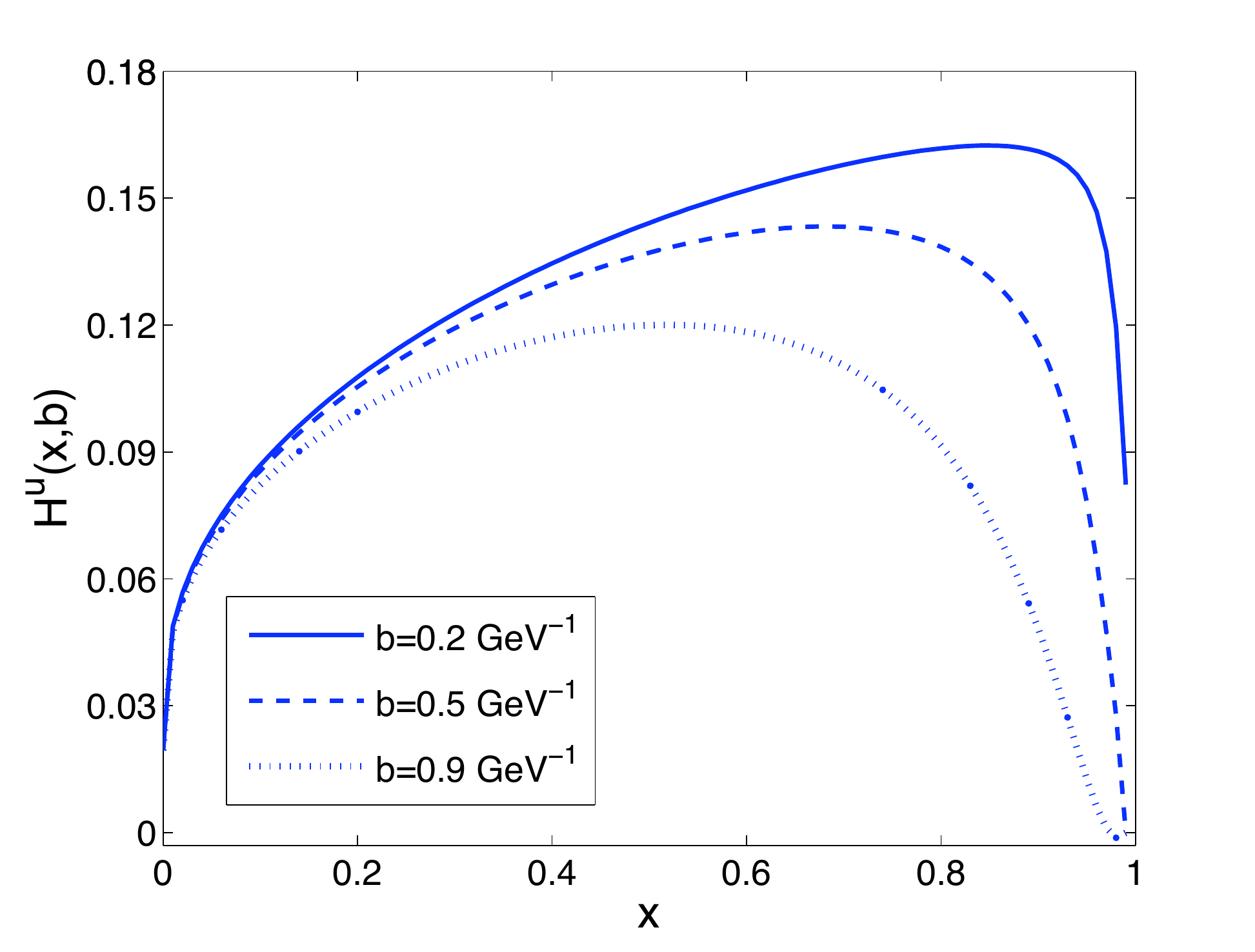}
\hspace{0.1cm}%
\small{(b)}\includegraphics[width=7cm,height=6cm,clip]{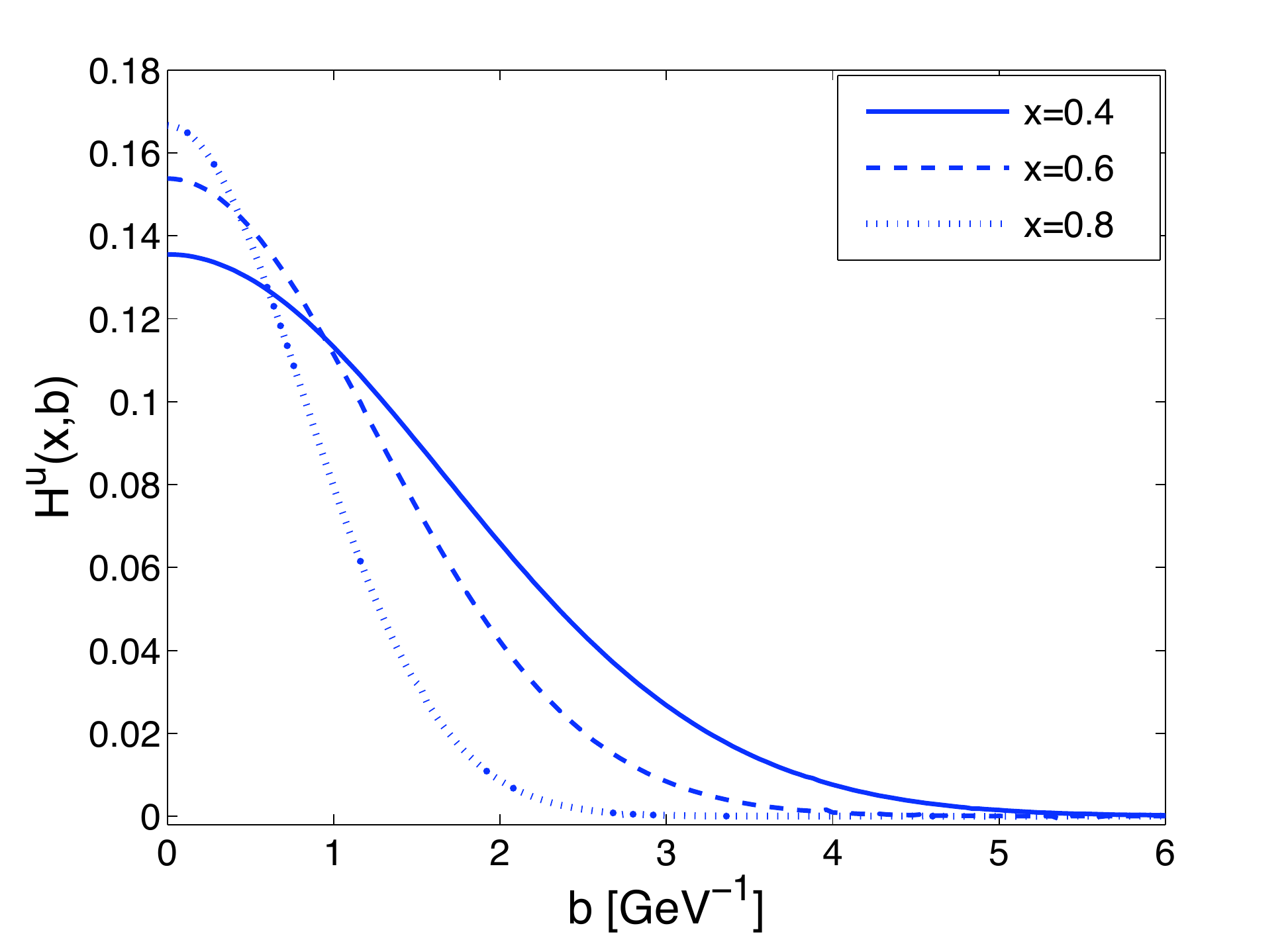}
\end{minipage}
\begin{minipage}[c]{0.98\textwidth}
\small{(c)}\includegraphics[width=7cm,height=6cm,clip]{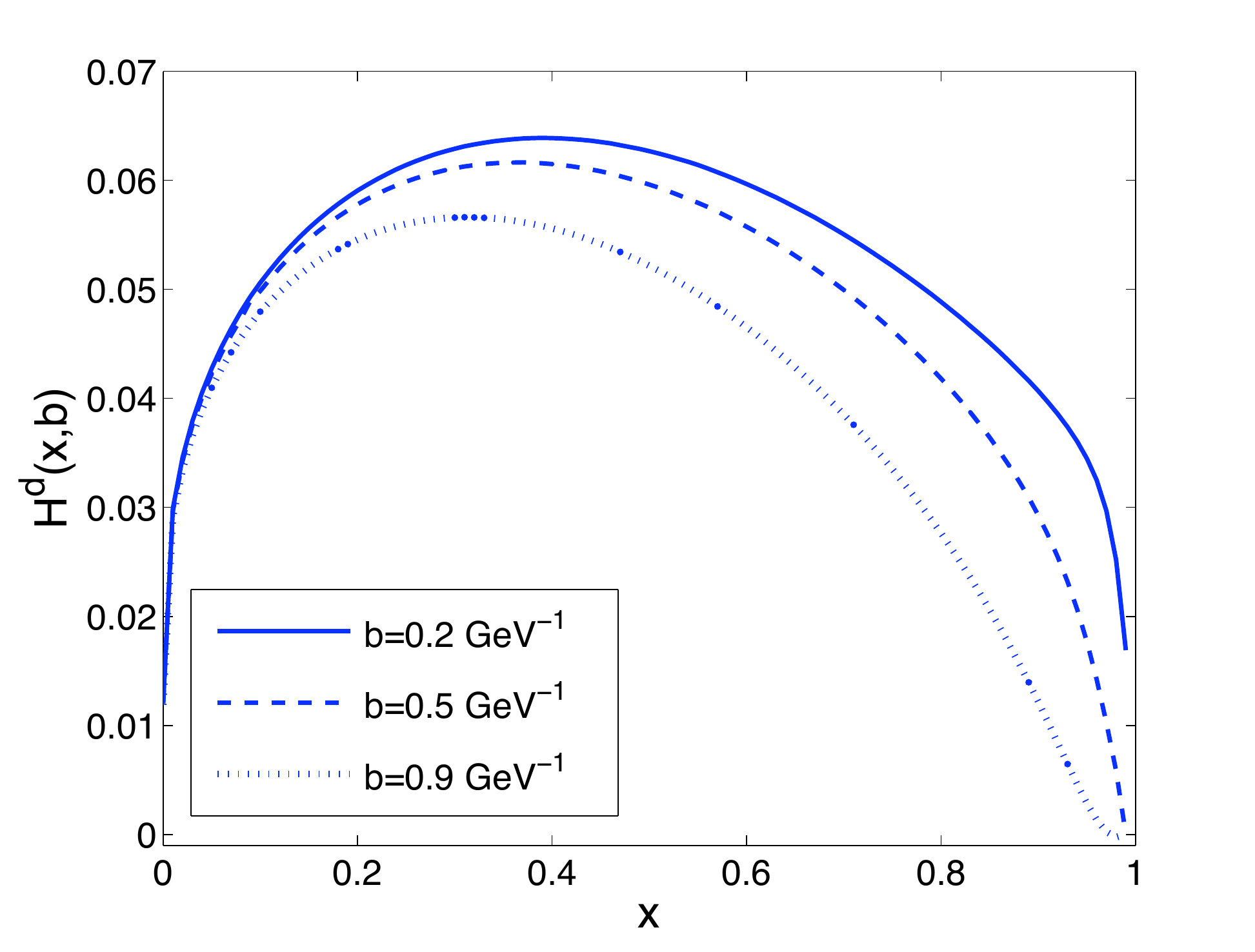}
\hspace{0.1cm}%
\small{(d)}\includegraphics[width=7cm,height=6cm,clip]{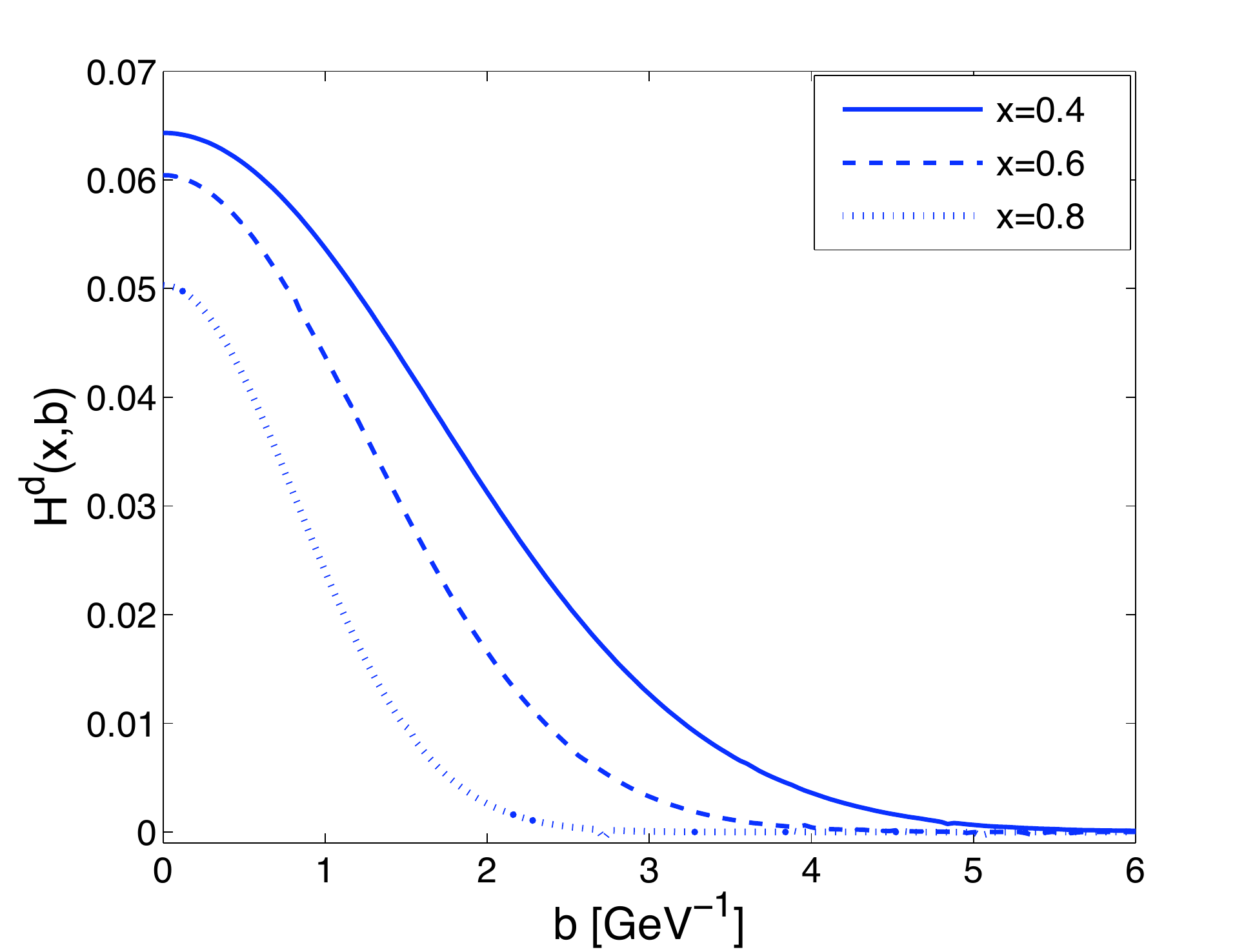}
\end{minipage}
\caption{\label{H_b_x}(Color online) Plots of (a) $H^u(x,b)$ vs $x$ for fixed values of impact parameter $b = \mid b_\perp \mid$;    (b) $H^u(x,b)$ vs $b$ for fixed $x$;  (c) and (d) are the same as in (a) and (b)  but for $d$ quarks.}
\end{figure}

\begin{figure}[htp]
\begin{minipage}[c]{0.98\textwidth}
\small{(a)}\includegraphics[width=7cm,height=6cm,clip]{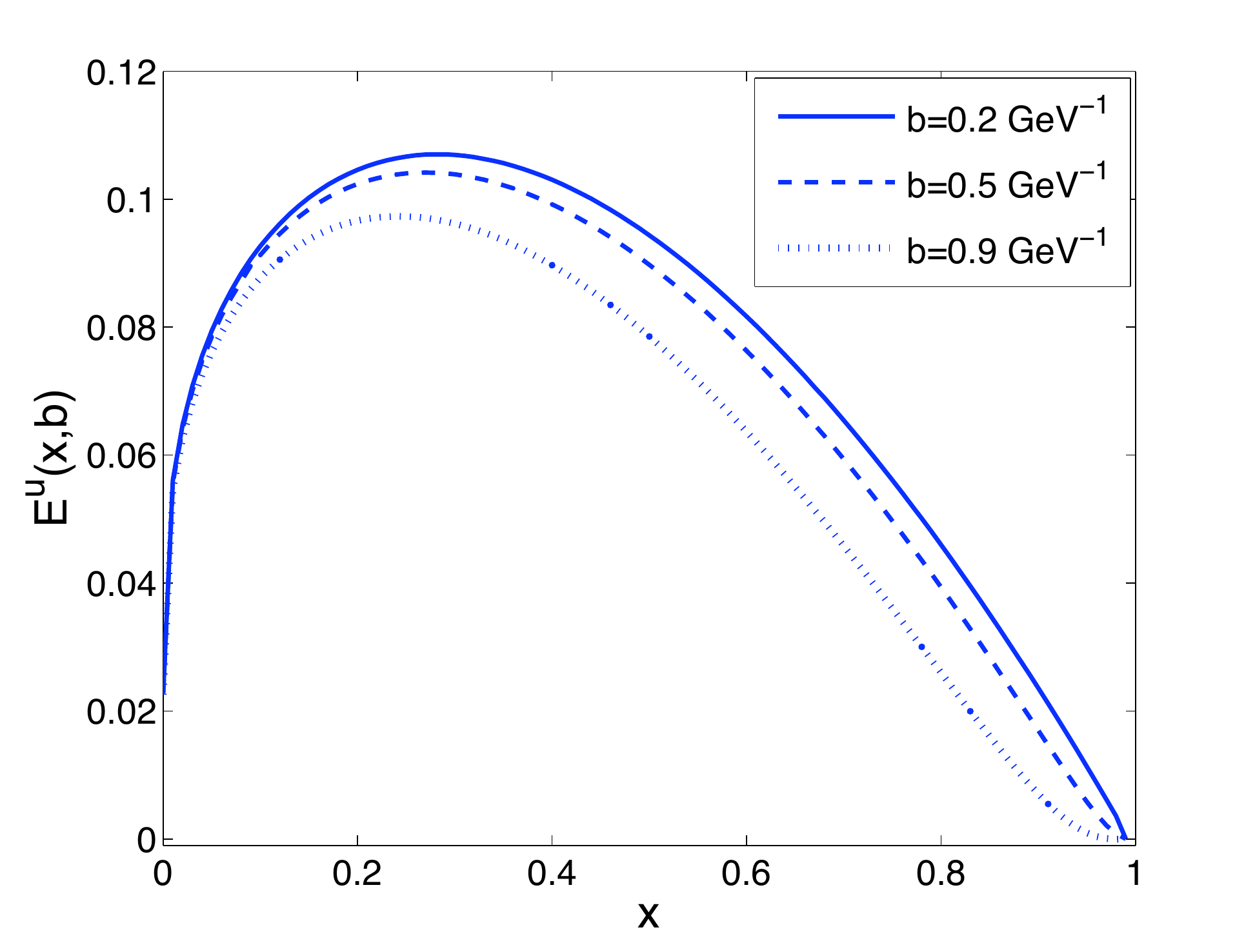}
\hspace{0.1cm}%
\small{(b)}\includegraphics[width=7cm,height=6cm,clip]{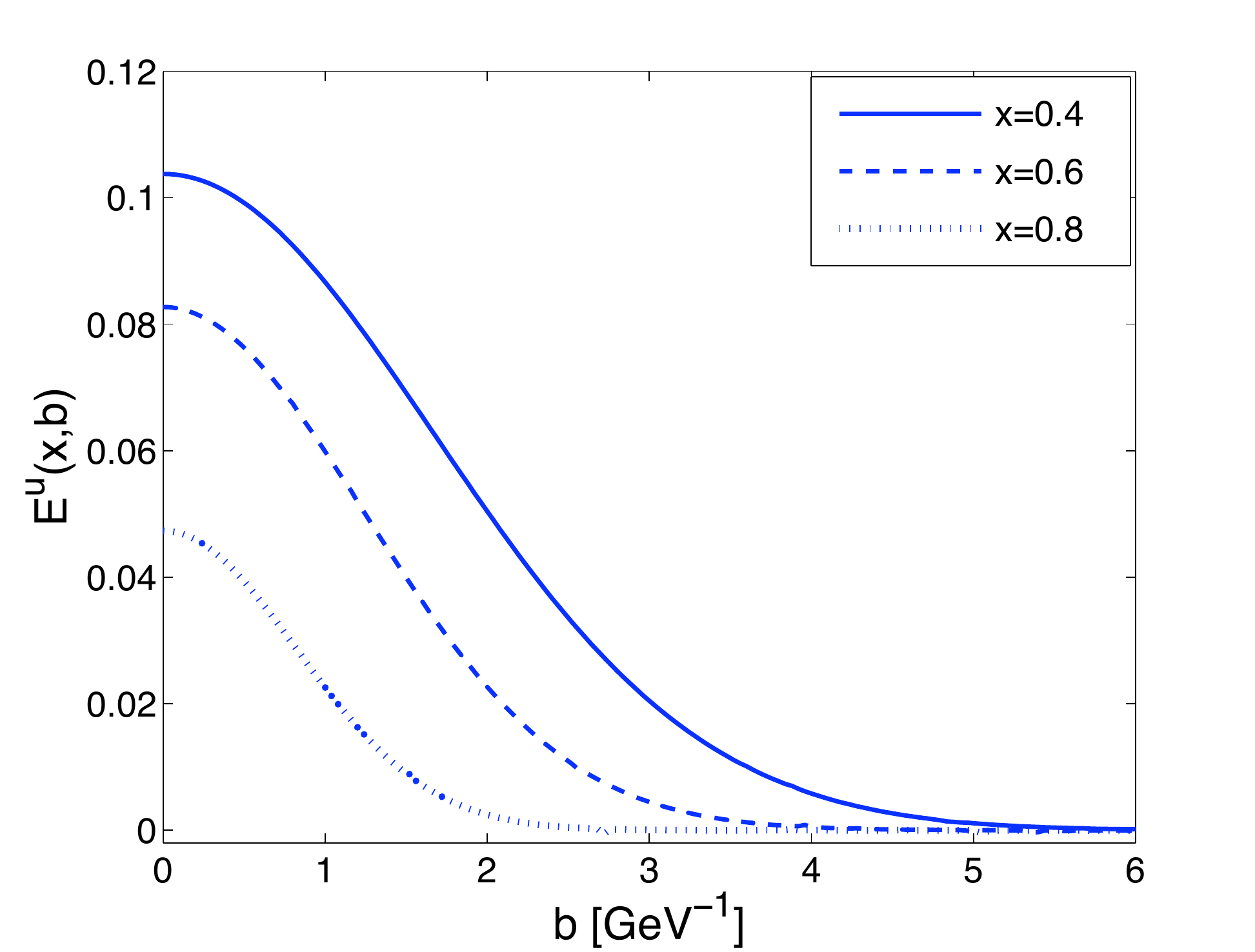}
\end{minipage}
\begin{minipage}[c]{0.98\textwidth}
\small{(c)}\includegraphics[width=7cm,height=6cm,clip]{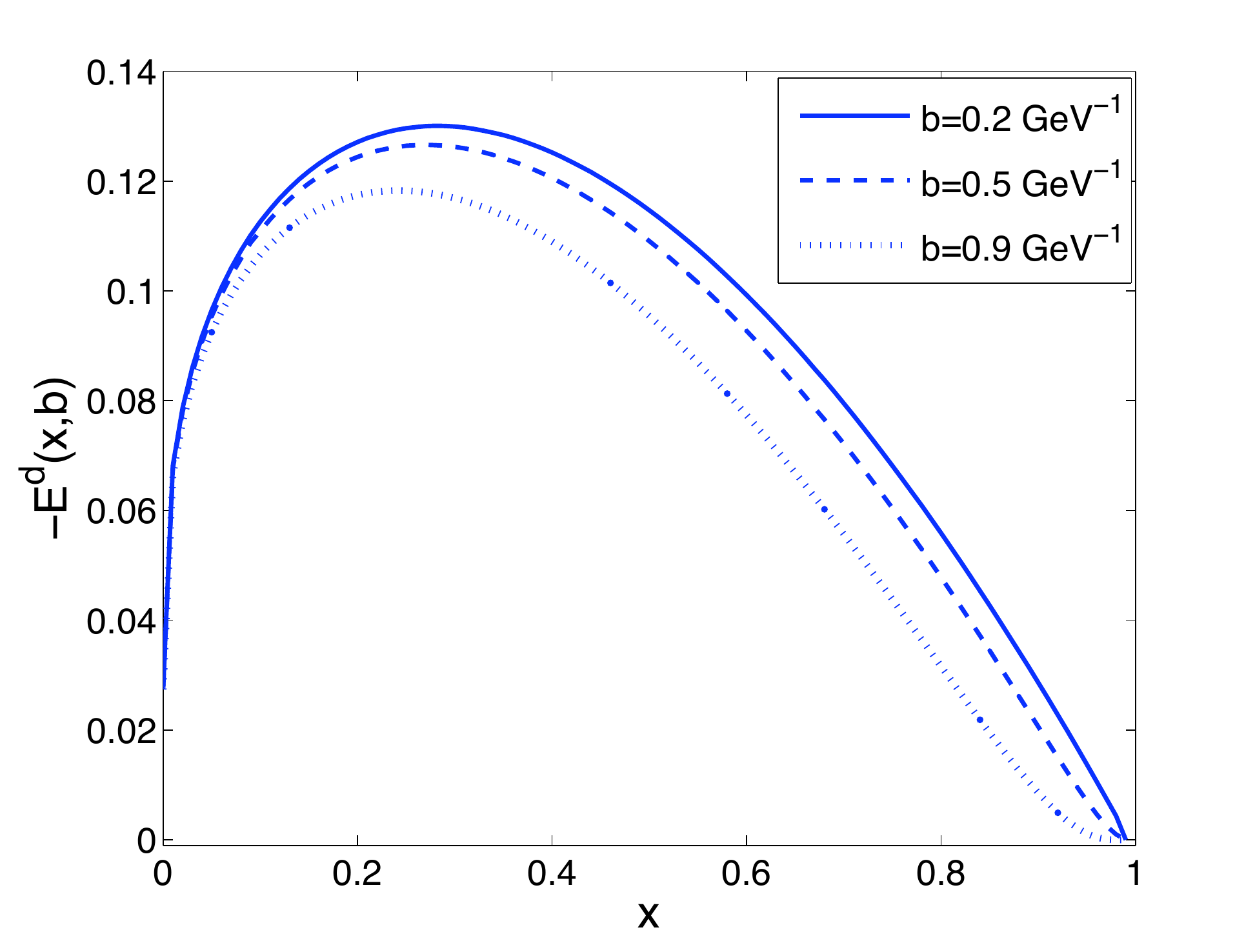}
\hspace{0.1cm}%
\small{(d)}\includegraphics[width=7cm,height=6cm,clip]{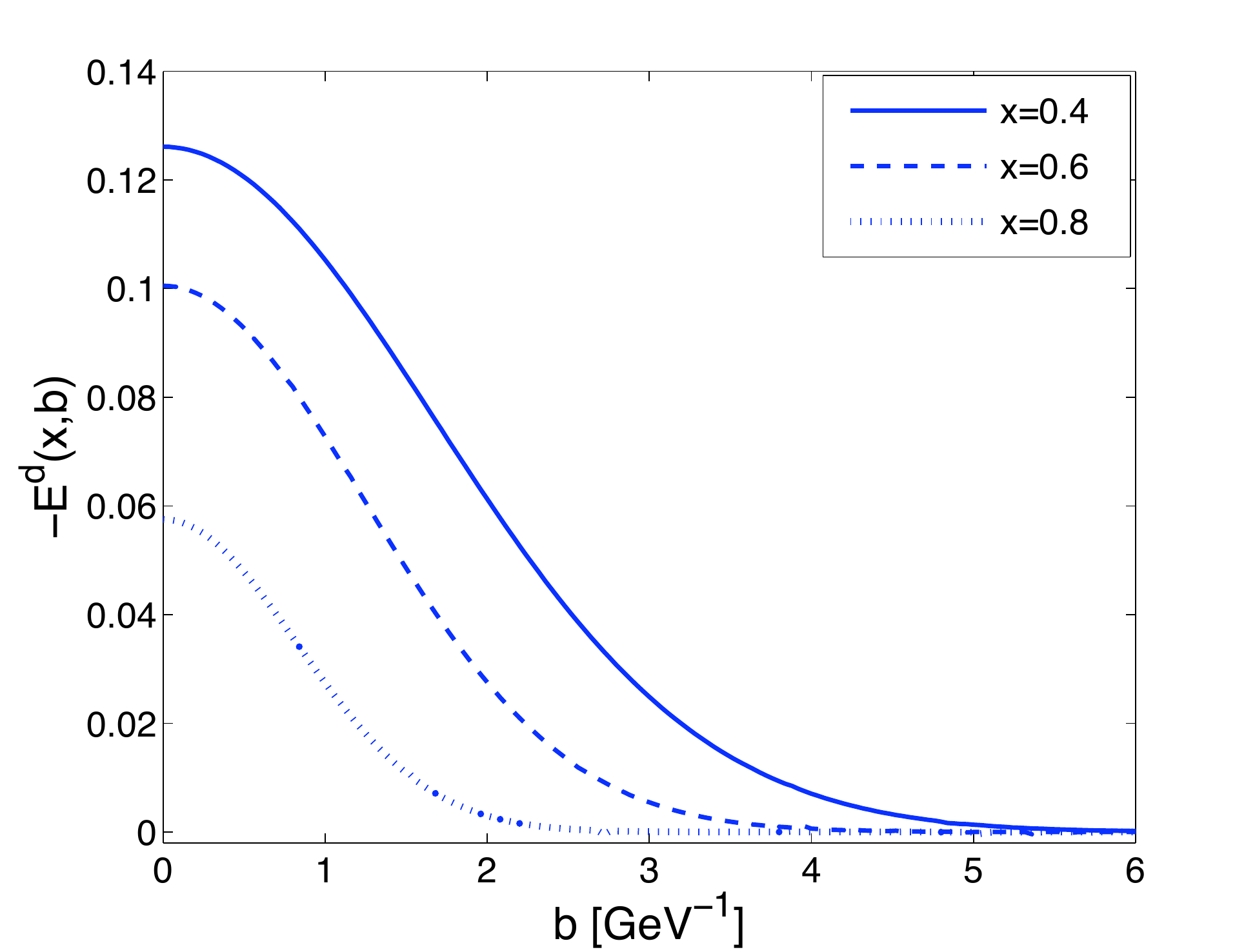}
\end{minipage}
\caption{\label{E_b_x}(Color online) Plots of (a) $E^u(x,b)$ vs $x$ for fixed values of  $b  = \mid b_\perp \mid$; (b)  $-E^d(x,b)$ vs$x$ for fixed values of  $b  = \mid b_\perp \mid$ for $d$ quarks.}
\end{figure}

The transverse impact parameter  $b= \mid b_\perp \mid $ is a measure of  the 
transverse distance between
the struck parton and the center of momentum of the hadron and  satisfies  $\sum_i x_i b_i=0$, where the sum is over the number of partons. An estimate of the size of the bound state can  be obtained from the relative distance 
 between the struck parton and the center of momentum of the
spectator system and is given by  ${b\over 1-x}$ \cite{diehl}. However, as the spatial extension of the spectator system is not available from the GPDs, exact estimation of the nuclear size is not possible. 
In Fig.\ref{H_b_x}(a), we show the behavior of $H^{u}(x,b)$ in $x$ for  fixed values of the  impact parameter 
$b$ and in Fig.\ref{H_b_x}(b) we  show the same GPD as function of $b$ for fixed values of $x$, and the similar plots for  $d$ quark are  shown in Figs.\ref{H_b_x}(c) and (d). The GPDs
$E^{u/d}(x,b)$ are shown in   Fig.\ref{E_b_x}.

 The GPDs evaluated in this model are qualitatively similar to the GPDs calculated in AdS/QCD using another parametrization\cite{vega}. So,  it is 
  interesting to compare with a phenomenological model.  Here we compare the overall behaviors of the  GPDs in the impact parameter space with those obtained from a phenomenological model for  the proton proposed in \cite{Ahmed}. The GPDs in this model are given by
\be
H^q(x,t)&=& G^{\lambda^q}_{M^q_x}(x,t) x^{-\alpha^q-\beta^q_1(1-x)p_1t},\\
E^q(x,t)&= &\kappa_q G^{\lambda^q}_{M^q_x}(x,t) x^{-\alpha^q-\beta^q_2(1-x)p_2t},
\ee
 where the first part is derived from the spectator model and modified by the Regge term to have proper behavior at low $x$.  The parameters are fixed by fitting the form factors. The details of the functional forms and the values of the parameters can be found in \cite{Ahmed}.    The impact parameter dependent GPDs from this model have been studied in \cite{CMM}.  One should remember  that the  valence GPDs we have considered here in AdS/QCD  are not exactly the same as GPDs in this model and so exact agreement is not expected, but it is interesting to compare and contrast the overall behaviors of the GPDs from these two models as we expect that the valence GPDs dominate the overall behavior for the GPDs in a proton.

\begin{figure}[htbp]
\begin{minipage}[c]{0.98\textwidth}
\small{(a)}
\includegraphics[width=7cm,height=6cm,clip]{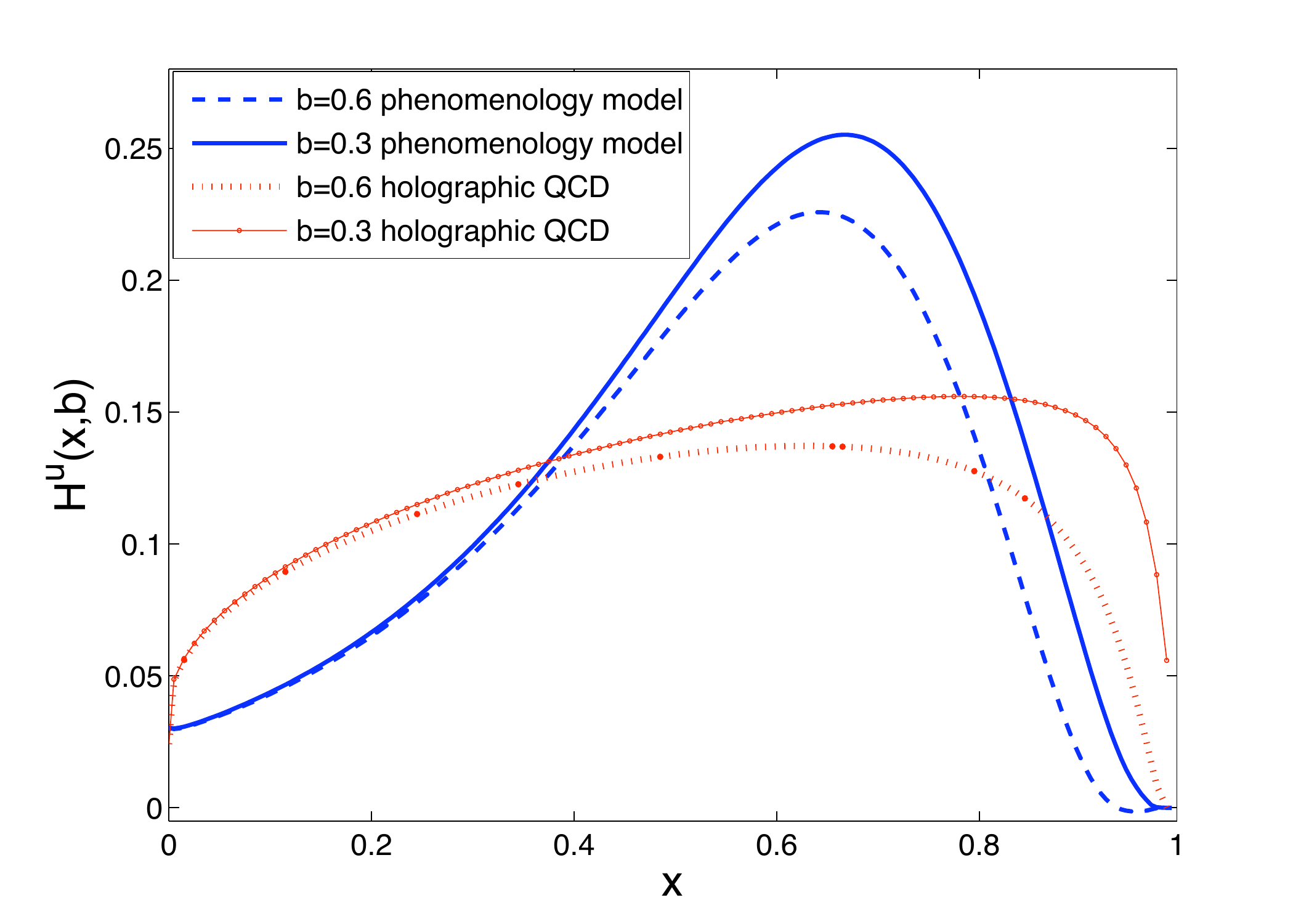}
\hspace{0.1cm}%
\small{(b)}\includegraphics[width=7cm,height=6cm,clip]{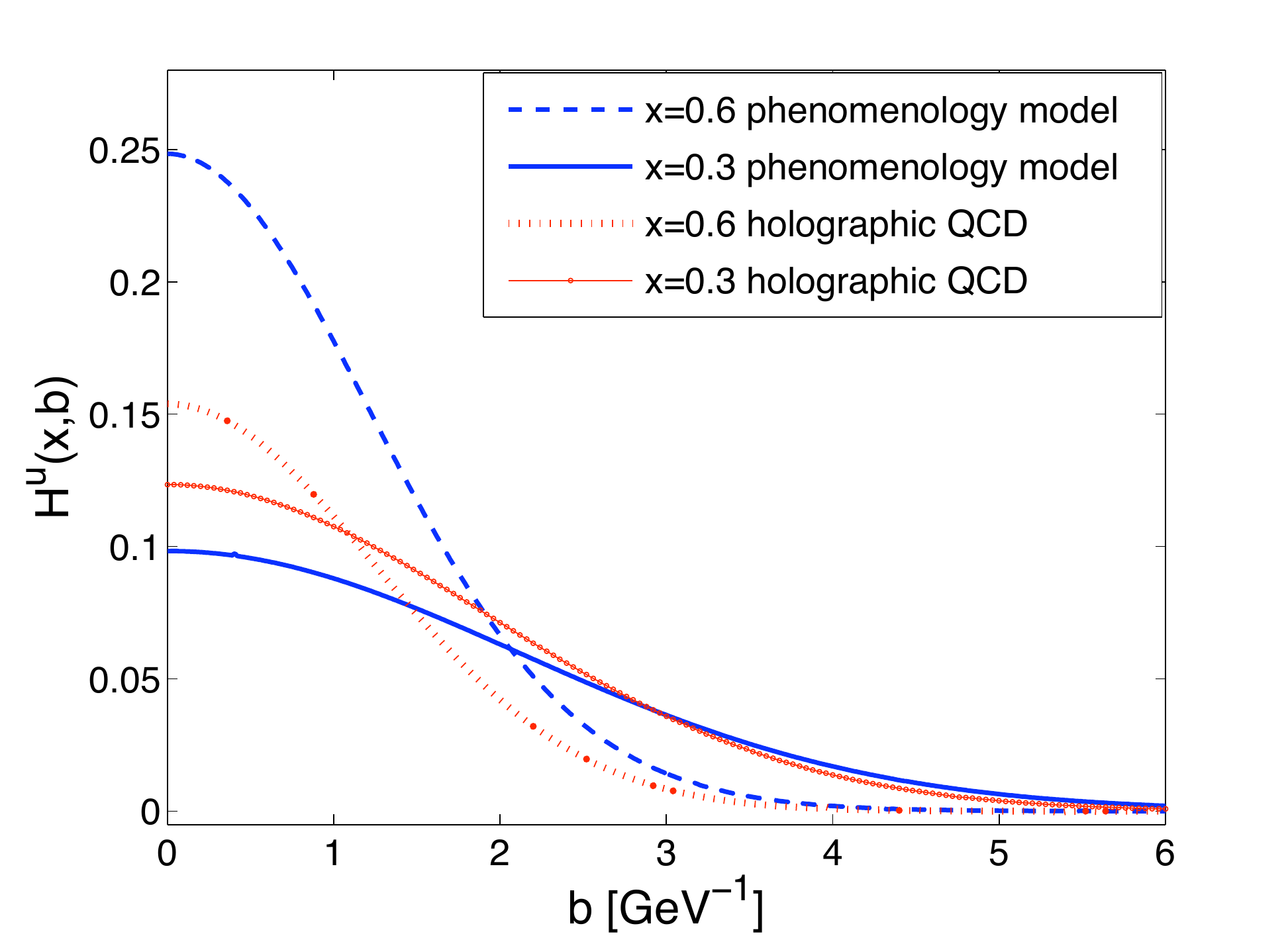}
\end{minipage}
\begin{minipage}[c]{0.98\textwidth}
\small{(c)}\includegraphics[width=7cm,height=6cm,clip]{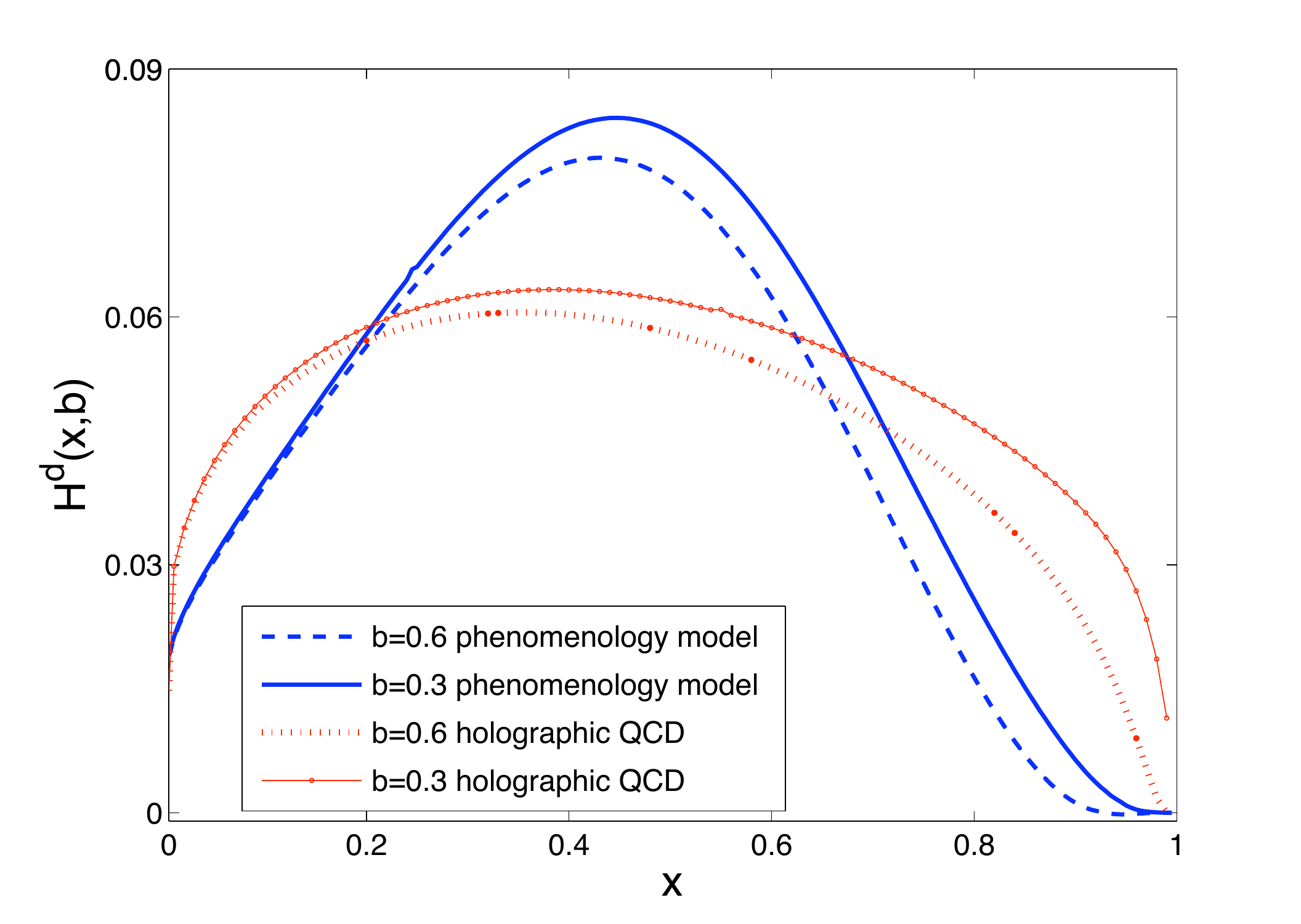}
\hspace{0.1cm}%
\small{(d)}\includegraphics[width=7cm,height=6cm,clip]{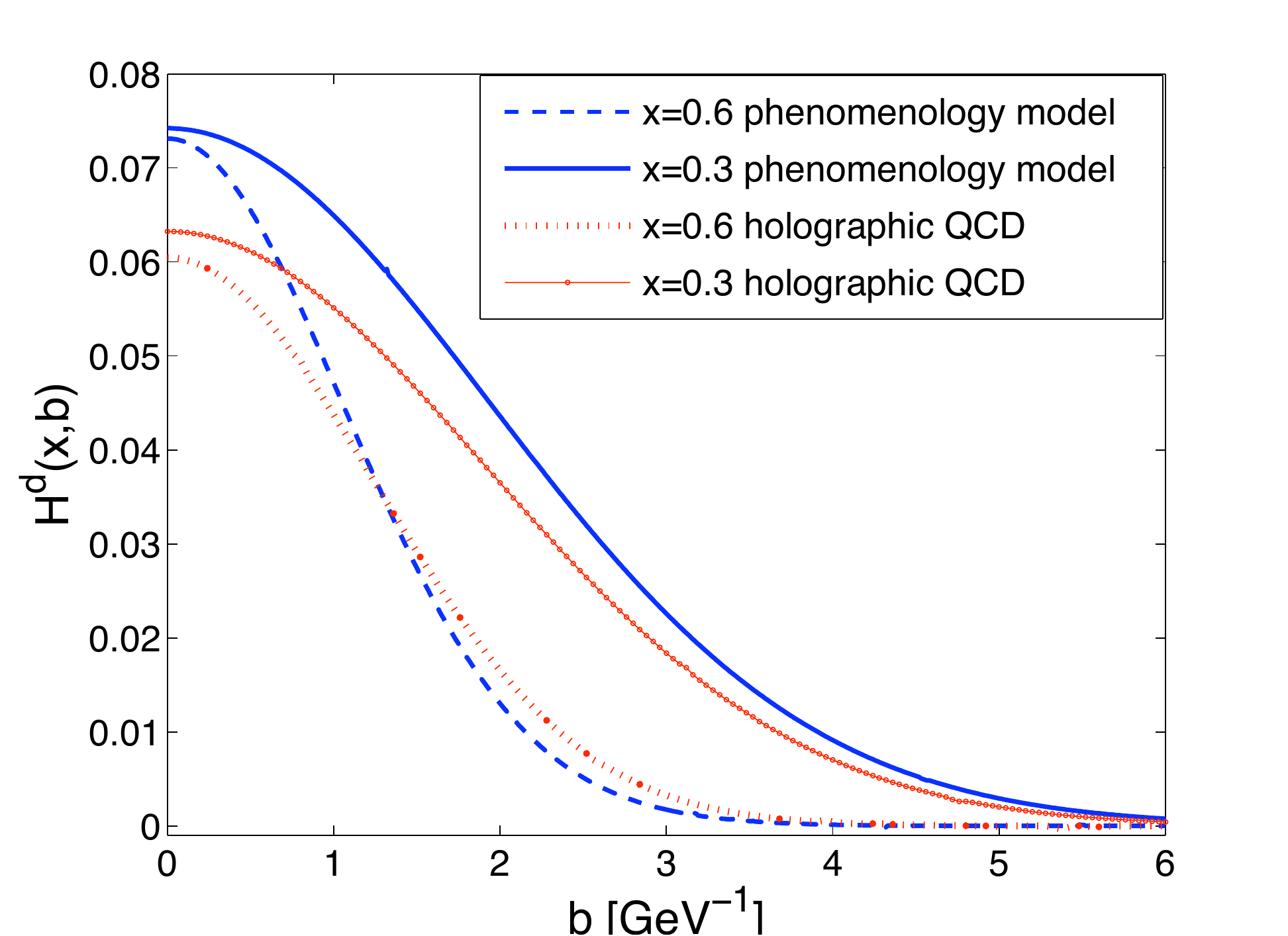}
\end{minipage}
\caption{\label{H_models}(Color online) Plots of (a) $H^u(x,b)$ vs $x$ for fixed values of impact parameter $b = \mid b_\perp \mid$   (b) $H^u(x,b)$ vs $b$ for fixed values of $x$,  (c) same as in (a) but for $d$ quark and (d) same as in (b) but for $d$-quark.}
\end{figure}
\begin{figure}[htpb]
\begin{minipage}[c]{0.98\textwidth}
\small{(a)}\includegraphics[width=7cm,height=6cm,clip]{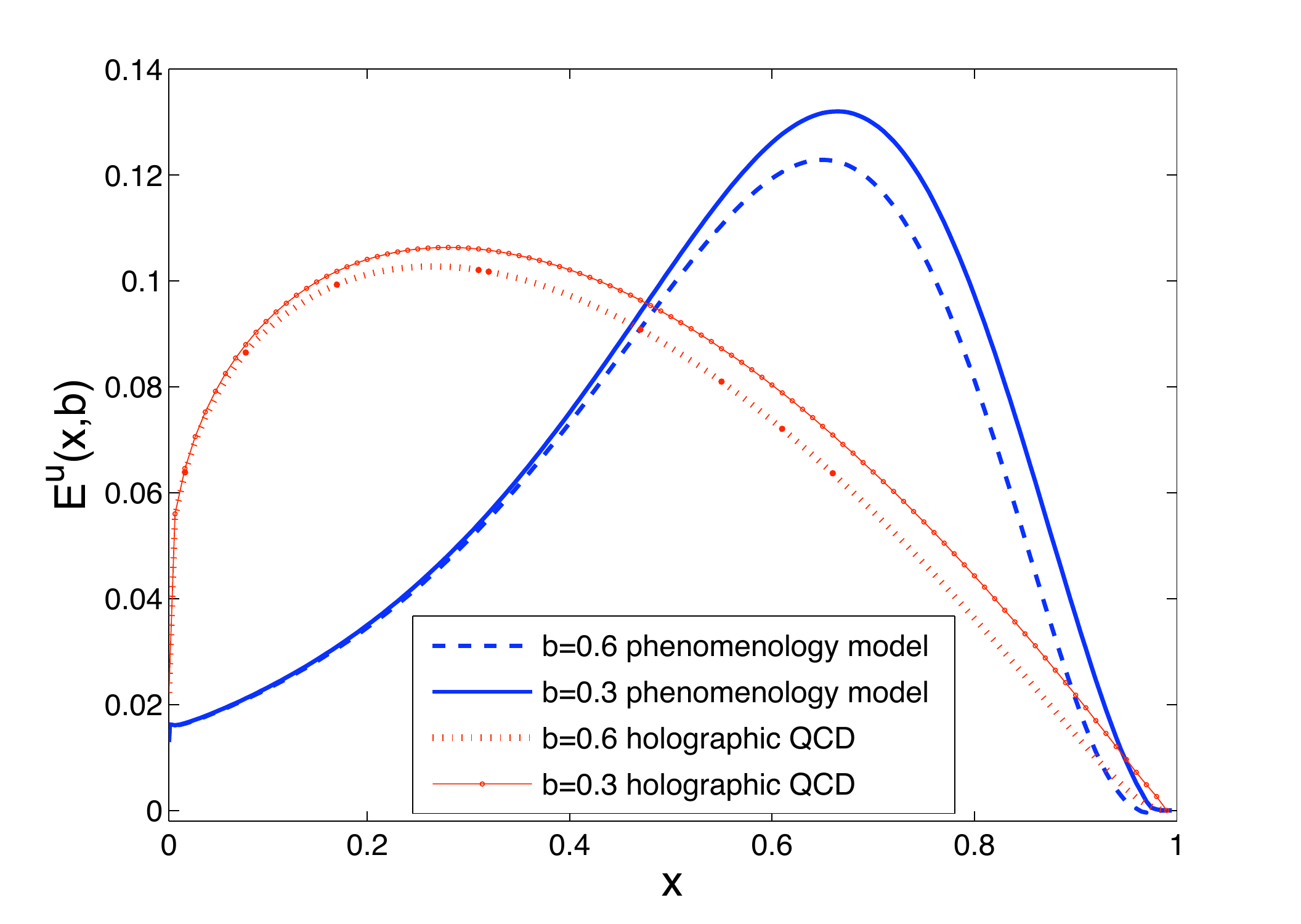}
\hspace{0.1cm}%
\small{(b)}\includegraphics[width=7cm,height=6cm,clip]{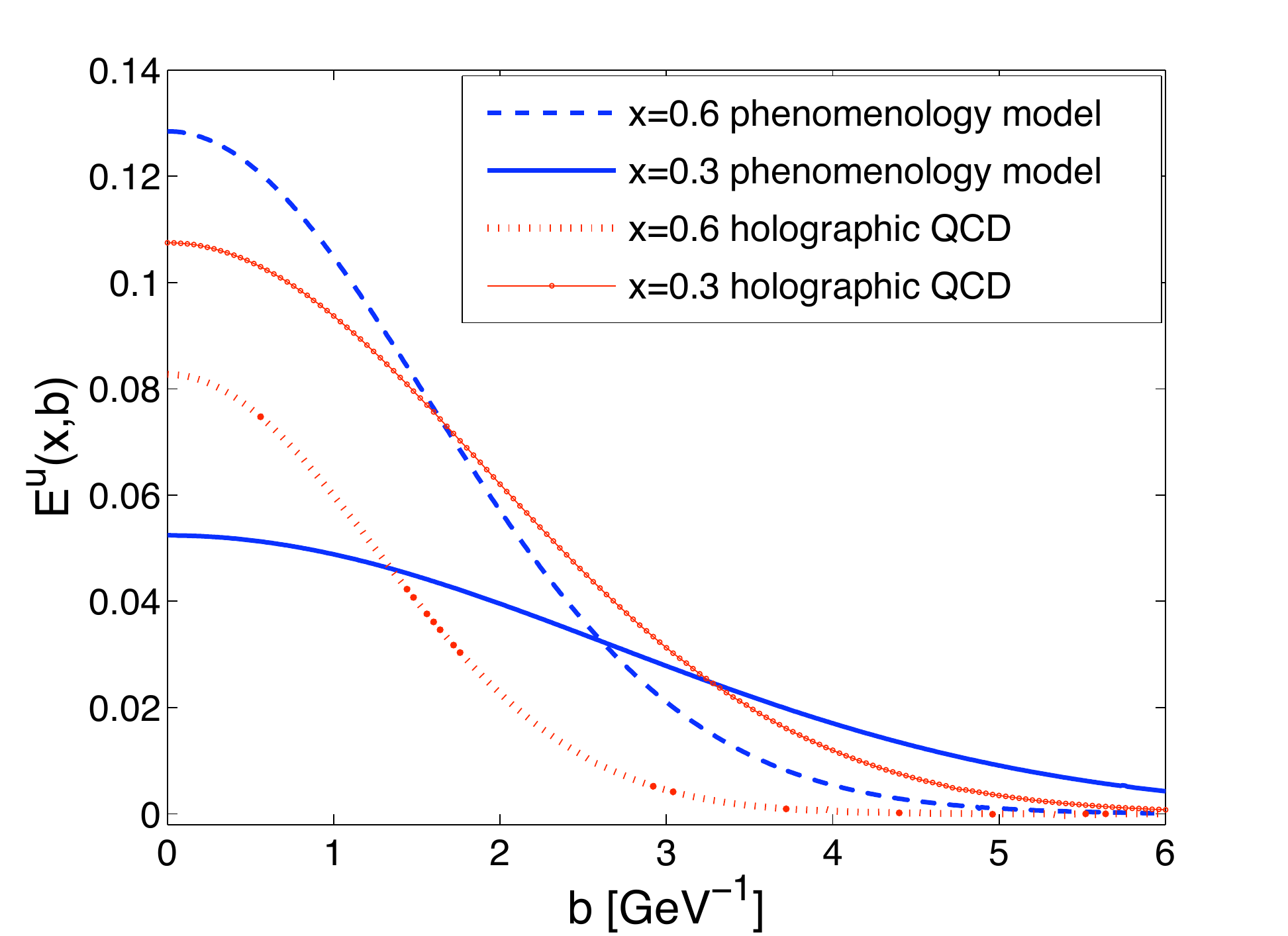}
\end{minipage}
\begin{minipage}[c]{0.98\textwidth}
\small{(c)}\includegraphics[width=7cm,height=6cm,clip]{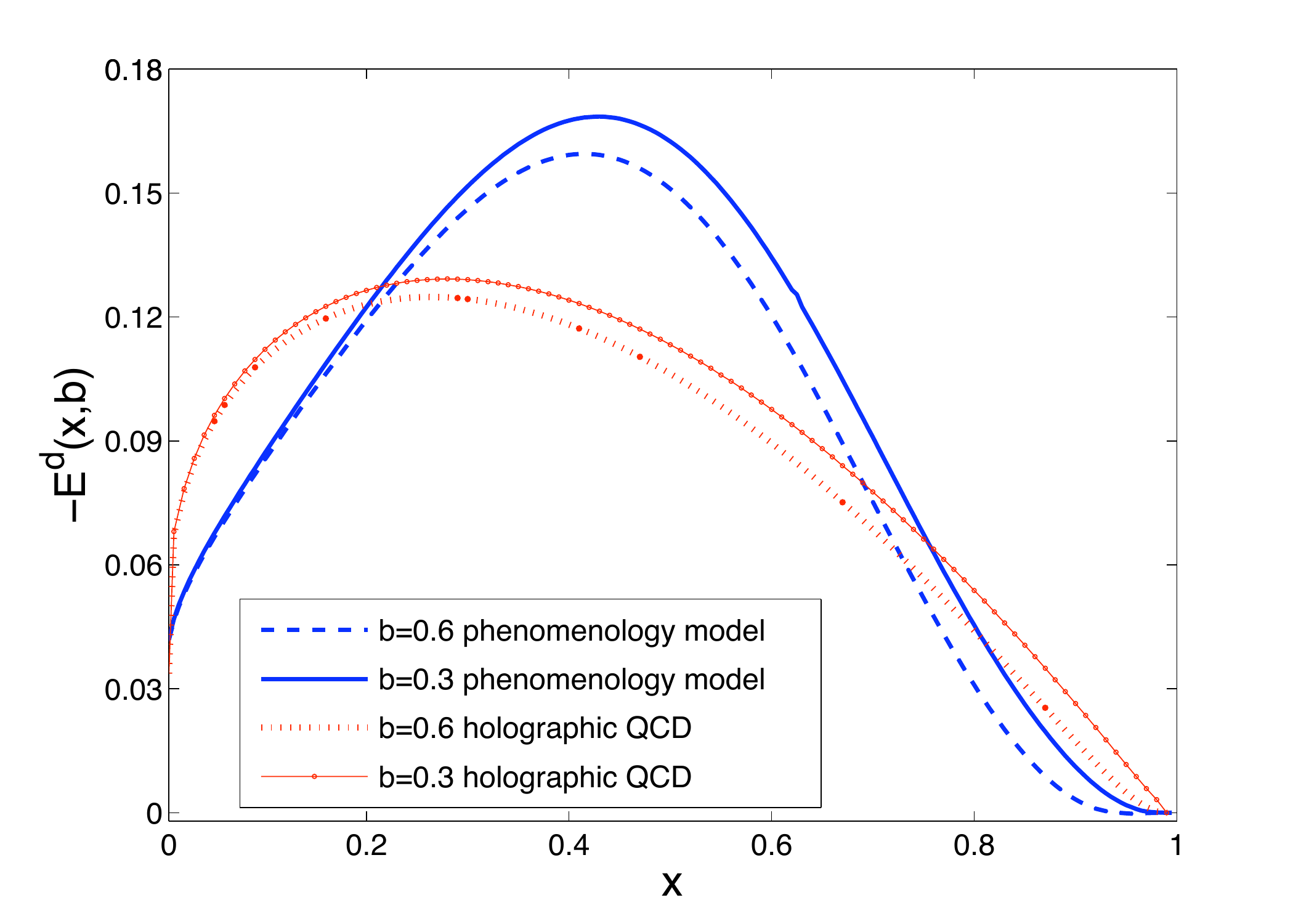}
\hspace{0.1cm}%
\small{(b)}\includegraphics[width=7cm,height=6cm,clip]{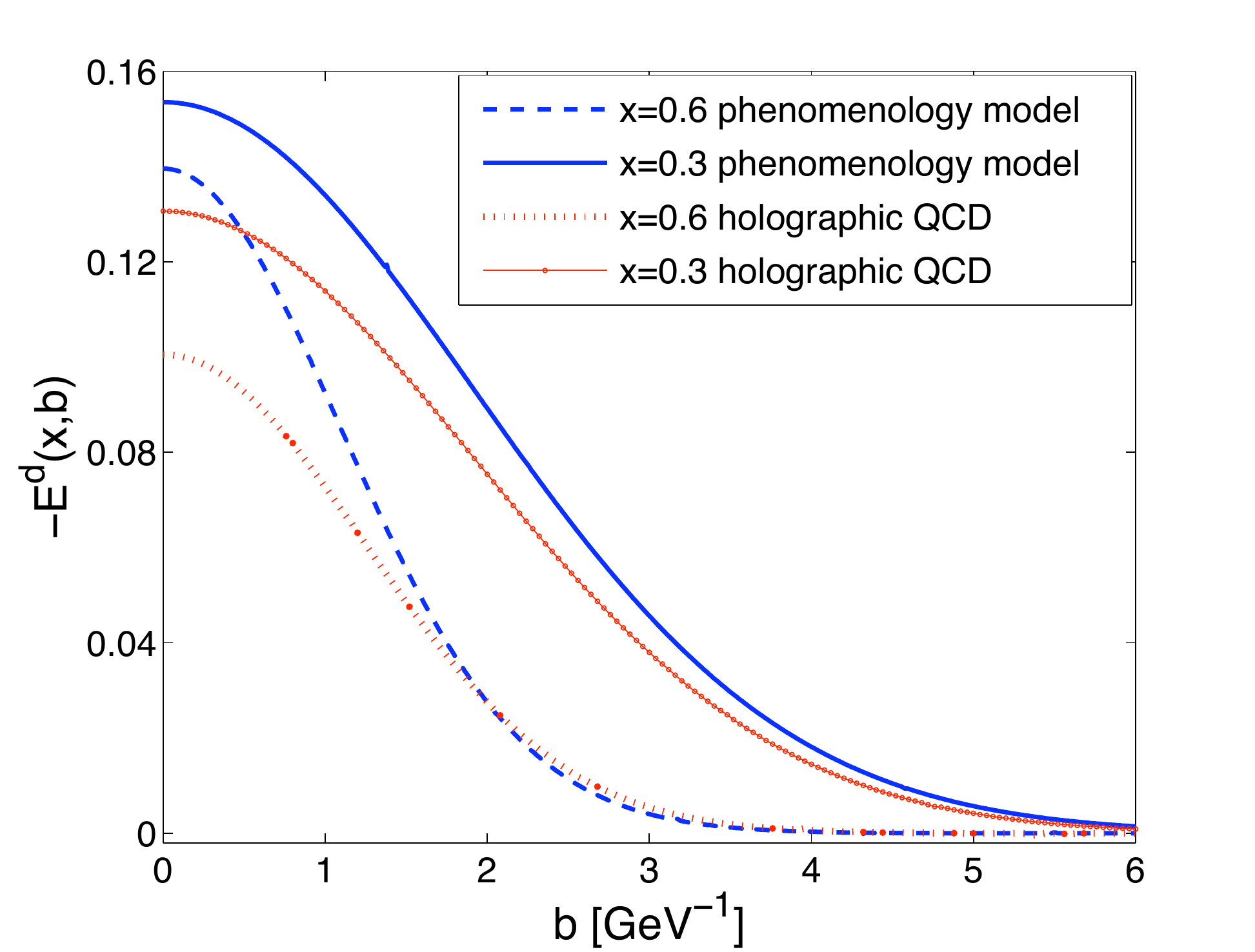}
\end{minipage}
\caption{\label{E_models}(Color online) 
 Plots of (a) $E^u(x,b)$ vs $x$ for fixed values of  $b  = \mid b_\perp \mid$;
(b) $E^u(x,b)$ vs. $b$ for fixed values of $x$;
 (c) the same as in (a) but  for $d$-quarks; and (d)  the same as in (b) but for $d$-quarks.}
 \end{figure}

In Fig.\ref{H_models}, we  compare the impact parameter dependent  proton GPD $H(x,b)$   from AdS/QCD with  the model mentioned above, for both $u$ and $d$ quarks. The GPDs fall off slowly for large $x$ in  AdS/QCD compared to the model, while in the impact parameter space they look almost the same except for  the difference in the magnitudes. In Fig.\ref{E_models} we have compared the two models for the proton GPD $E(x,b)$.  The behavior in $x$ for $u$-quarks is quite different in the two models, while they agree better for $d$-quarks, and again the  GPDs from AdS/QCD fall off slowly at large $x$ compared to the other model.  In the model, the behavior of $E(x,b)$ for $u$ and $d$ quarks is quite different when plotted against $x$ for fixed values of impact parameter $b$, whereas in 
 AdS/QCD, it shows almost the same behavior for both $u$ and $d$ quarks. 
 The GPD $E(x,b)$ in both models  agrees better  in impact parameter space  for  the $d$-quark than the $u$-quark.  The difference  with the AdS/QCD results can be attributed to the fact that AdS/QCD provides only the valence wave functions in the semiclassical approximation.
 It is interesting to note that  in both cases, at small values of impact parameter $b$, the  GPD $H(x,b)$ is larger for  $u$-quarks than  $d$-quarks whereas the magnitude of  the GPD $E(x,b)$ is marginally  larger for $d$-quarks than the that for $u$-quarks.  Thus it  is interesting to check with other models whether this is  a model independent result.
 
 \vskip0.2in
\noindent
{\bf Concluding remarks}\\
 In this work we have evaluated the GPDs for protons using the LFWFs obtained from AdS/QCD.     It is shown \cite{BT2} that the electromagnetic form factors for protons and neutrons calculated by using the AdS/QCD wave functions fit well with the experimental results.   The GPDs are calculated using the form factors and exploiting the integral representation of the bulk-to-boundary propagator in AdS space.  The valence GPDs thus obtained in the impact parameter space 
 have been compared  with the GPDs obtained from a phenomenological model. Though  in the AdS/QCD we have only valence GPDs  it is interesting to note that their behaviors are quite similar and agree well in impact parameter space with a phenomenological model for GPDs.
  But  variation of the GPDs from AdS/QCD are  slower  than the other model when compared to the behaviors in $x$ for both $u$ and $d$ quarks. 


\vskip0.2in
\noindent
{\bf Acknowledgements:} The authors acknowledge useful discussions with  Stan Brodsky and  thank  Stan Brodsky,  Guy de T\'eramand  and Valery Lyubovitskij for going through  the previous version of the manuscript and for their  valuable comments and references on the  AdS/QCD and its applications.



\end{document}